\documentclass{article}
\usepackage{alife10}
\usepackage{verbatim}

\title{Strategies for fast convergence in semiotic dynamics}
\author{Andrea Baronchelli$^{1}$, Luca Dall'Asta$^{2}$, Alain Barrat$^{2}$ \and
  Vittorio Loreto$^{1}$ \\
\mbox{}
$^1$Dipartimento di Fisica and SMC center, Universit\`a ``la
  Sapienza'', Roma, Italy \\
$^2$ Laboratoire de Physique Th\'eorique
(UMR du CNRS 8627), Universit\'e de Paris-Sud, France \\ 
Andrea.Baronchelli@roma1.infn.it}

\begin{document}
\maketitle

\begin{abstract}
  Semiotic dynamics is a novel field that studies how semiotic conventions
  spread and stabilize in a population of agents. This is a central issue both
  for theoretical and technological reasons since large system made up of
  communicating agents, like web communities or artificial embodied agents
  teams, are getting widespread. In this paper we discuss a recently
  introduced simple multi-agent model which is able to account for the
  emergence of a shared vocabulary in a population of agents. In particular we
  introduce a new deterministic agents' playing strategy that strongly
  improves the performance of the game in terms of faster convergence and
  reduced cognitive effort for the agents.
\end{abstract}

\section{Introduction}

Imagine a population of artificial embodied agents exploring an unknown
environment. One of the first tasks they should face is exchanging
informations about their discoveries. In particular, when a new 'thing' is
met, they should be able to agree on its identification. If the agents were
endowed with short range communication systems only, the agreement would take
place locally, and, in a second time, should hopefully become global. But how
could this happen? A first hypothesis is that they could agree on the
geographical location of the object and everybody could go and see it. This is
of course very costly, and definitely not efficient since each new finding of
the same object would require the same spending procedure. Moreover it would
require a mechanism of global coordination, which is not always available.
Instead, it would be desirable that each agent could recognize the object the
first time it sees it and then assign a true 'name' to it. The global
agreement on the name would then allow for a great saving of time and would
also be crucial for the birth of a communication system among agents. This
kind of problems inspired the well known "Talking Heads Experiment" in which
embodied software agents were shown to be able of bootstrapping a
shared set of semiotic conventions~\cite{talking_heads}.

In the following, we shall assume that our agents are endowed with
the necessary tools 
needed to recognize and physically deal with an object and we shall
concentrate on the dynamics that leads to the obtention of a shared set of
conventions in a population. This is the general problem investigated by the
field of Semiotic Dynamics, according to which language is an evolving and
self organized system~\cite{steels_dynsys}. The term evolving must not be
misleading. The evolution of language across generations is a well
investigated aspect, and multi-agent modeling has already proved to be a
powerful tool for its investigation~\cite{hurford89,nowak99}. In our context,
on the other hand, the focus is on much shorter timescales, so that
we are not dealing with 
transmission across generations nor, more in general, Darwinian concepts. 
The issue of the self-organization of language on
fast temporal scales is, of course, of the outmost importance and generality.
Besides artificial systems, where it is obvious that the agreement has to take
place rapidly, it concerns human dynamics, too. In particular the user based
tagging systems presently spreading on the web (such as del.icio.us or
flickr.com), where users manage tags to share and categorize informations,
offer wonderful examples of self-organized communication systems. Gaining
hints into population-scale semiotic dynamics is then important for a twofold
reason. On the one hand it is necessary to interpret and understand presently
occurring phenomena, and on the other hand it can be very important to provide
indications for the design of large scale technological systems.

In this paper we focus on a recently introduced multi-agent
model~\cite{ng_first}, inspired by the so-called Naming Game~\cite{ng1995},
which, though being very simple in its definition, is able to account for the
birth of a shared set of conventions in a population. We investigate also how
the properties of the system change with the population size. We then study
how a modification of the rules of the original model, from random to
deterministic strategies, allows to improve the performance of
the population: both the time required for the consensus to emerge 
and the agents memory requirements are indeed substantially lowered.

\section{A simple model of semiotic dynamics}

Let us consider a population of $N$ agents which perform pairwise games in
order to agree on the name to assign to a single object. Each agent is
characterized by his inventory or memory, i.e. a list of name-object
associations that is empty at the beginning of the process and evolves
dynamically during time. At each time step, two agents are randomly selected,
one to play as speaker, the other one as hearer, and interact according to the
following rules:

\begin{itemize}
\item The speaker has to transmit a name to the hearer. If his inventory is
  empty, he invents a new word, otherwise he selects \textit{randomly} one of
  the names he knows;
  
\item If the hearer has the uttered name in his inventory, the game is a
  \underline{success}, and both agents delete all their words but the winning
  one;
  
\item If the hearer does not know the uttered word, the game is a
  \underline{failure}, and the hearer inserts the word in its inventory, i.e.
 he learns it.
\end{itemize}

A remark concerning the presence of a single object is in order. From a
linguistic point of view it is equivalent to the rather strong assumption of
preventing homonymy, thus making different objects independent.  This
simplification allows for a strong reduction in the complexity of
the model and, moreover, does not seem so drastic when thinking of artificial
agents that assign randomly extracted real numbers to new objects.

Another point concerns the difference with some other well-known models of opinion
and consensus formations. In Axelrod's model~\cite{Axelrod:1997}, 
each agent is endowed with a vector of opinions, and can interact
with other agents only if their opinions are already close enough;
in Sznajd's model~\cite{Sznajd:2000} and in 
the Voter model~\cite{Krapivsky:1992}, the opinion can take only
two discrete values, and an agent takes deterministically the opinion
of one of his neighbors. Also in \cite{Deffuant:2000}, the opinion
is modeled as a unique variable and the evolution of two interacting agents
is deterministic. In the Naming Game model on the
opposite, each agent can potentially have an unlimited number of possible
discrete states (or words, names) {\em at the same time},  
accumulating in his memory different possible names
for the object: the agents are able to "wait" before reaching 
a decision. Moreover, each
dynamical step can be seen as a negotiation between speaker and
hearer, with a certain degree of stochasticity.

\begin{figure}[t]
\begin{center}
\includegraphics[width=7.5cm]{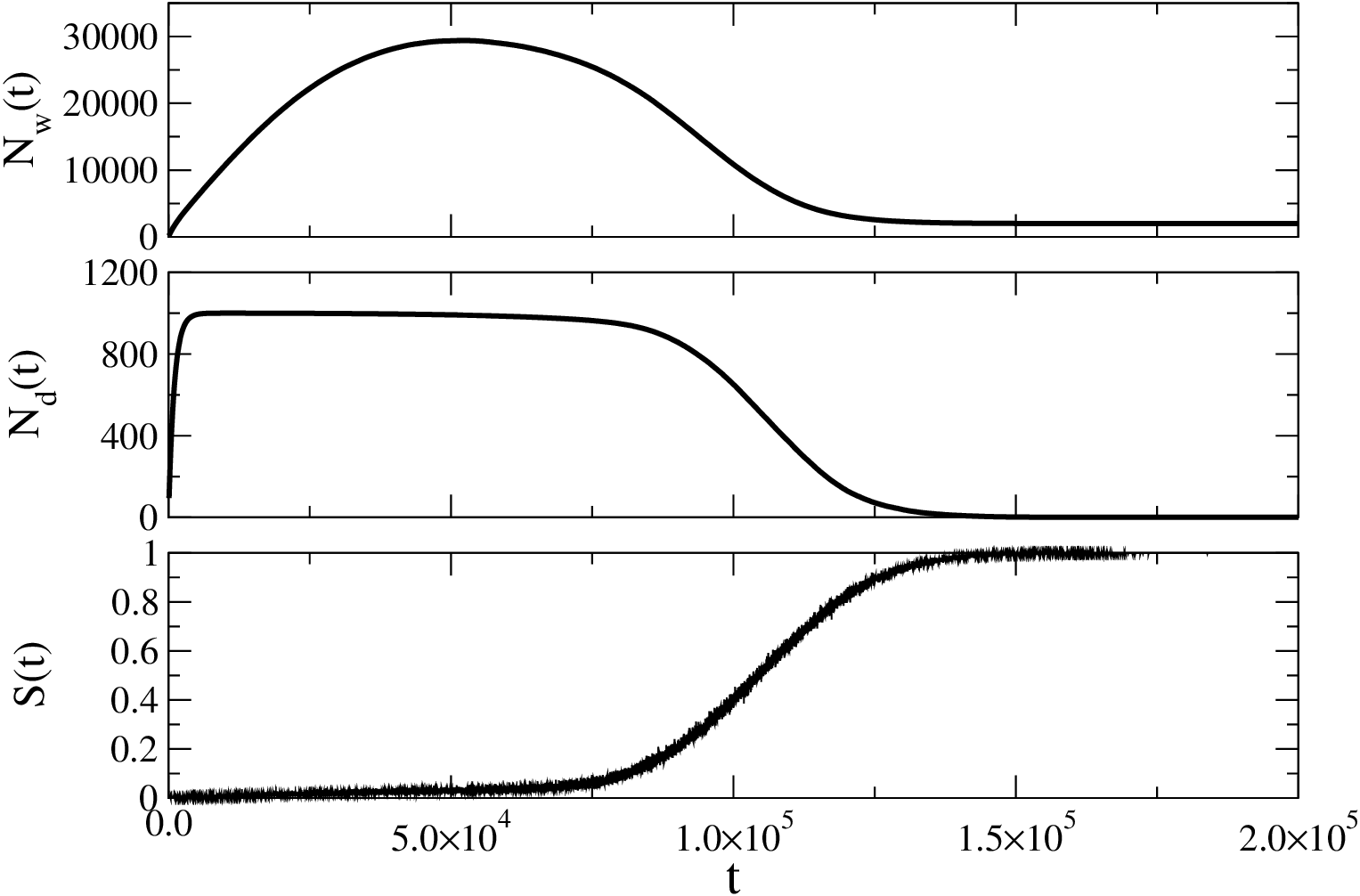}
\caption{Process evolution - The total number of words in the system,
  $N_w(t)$ (or total memory used), the number of different words, $N_d(t)$,
  and the success rate, $S(t)$, are plotted as a function of time. The final
  convergence state is characterized by the presence of the same unique word
  in the inventories of agents. Thus, at the end of the process, we have
  $N_w(t)=N$ and $N_d(t)=1$, while the probability of a success is equal to
  $1$ ($S(t)=1$). The curves have been obtained averaging over $300$ simulation
  runs for a population of $2\times 10^3$ agents.}
\label{f:classik2k}
\end{center}
\end{figure}

To understand the behavior of the system~(see
also~\cite{ng_first}), we report in Figure \ref{f:classik2k} three curves
obtained averaging several runs of the process in a population of
$N=2\times10^3$ agents. They represent the evolution in time of the total
number of words present in the system $N_w(t)$, which quantifies the total
amount of memory used by the process, of the number of different words,
$N_d(t)$, and of the success rate, $S(t)$, defined as the probability of a
successful interaction between two agents at time $t$. The first thing to be
noted is that the system reaches a final convergence state in
which all agents have the same unique word, i.e. a final proto-communication
system has been established. It is thus interesting to proceed
with a more detailed analysis of how this final state of global
communication emerges from purely binary interactions.

The process starts with a trivial phase in which the inventories are empty, so
that the invention process is dominating and $N/2$ different words are created
on average. This rapid transient is followed by a longer period of time in
which most interactions are unsuccessful ($S(t) \simeq 0$), and the sizes of
inventories keep growing. However the amount of memory used does not increase
indefinitely, since correlations are progressively built up among inventories 
and increase the probability of successful interactions. In particular, the
$N_w(t)$ curve exhibits a well identified peak, whose height and occurrence
time are important parameters to describe the process. Slightly after the peak,
there is a quite abrupt transition from a disordered state in which
communication among agents is difficult to a nearly optimal situation, which
is captured by a jump of the success rate curve. The process then ends
when the convergence state ($N_d(t)=1$, $N_w(t)=N$) is reached.
Finally, it is worth noting that the developed proto-communication system is
not only effective (each agent understands all the others), but also efficient
(no memory is wasted in the final state).

\begin{figure}[t]
\begin{center}
\includegraphics[width=7.5cm]{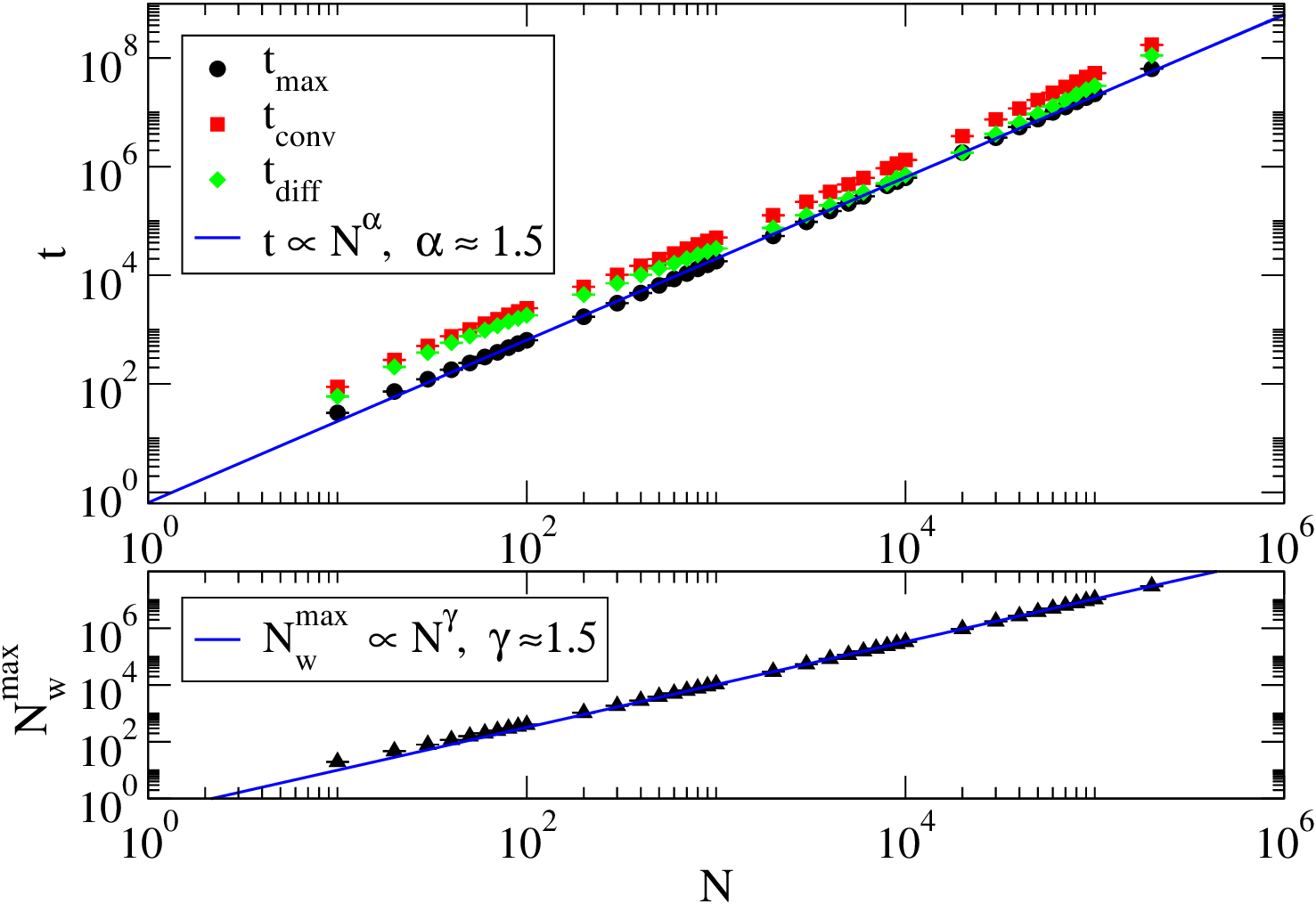}
\caption{Scaling with the population size $N$. In the upper graph 
  the scaling of the peak and convergence time, $t_{max}$ and $t_{conv}$, is
  reported, along with their difference, $t_{diff}$. All curves scale with the
  power law $N^{1.5}$. Note that $t_{conv}$ and $t_{diff}$ scaling curves
  present characteristic log-periodic oscillations.  The lower curve shows
  that the maximum number of words (peak height, $N_w^{max}=N_w(t_{max})$)
  obeys the same power law scaling.}
\label{f:scaling_mf}
\end{center}
\end{figure}

To gain a deeper comprehension of the process it is important to
investigate how the main features scale with the system size $N$. In
particular, it is relevant to know how the agents' cognitive effort in terms of
memory and the time required to reach the final state depend on the population
size. The global memory used is maximum when the number of
words is the highest, i.e. at the peak of the $N_w(t)$ curve. The
scaling of the peak time $t_{max}$ and height $N_w^{max}=N_w(t_{max})$
are therefore studied, together with the convergence time $t_{conv}$,
in Figure~\ref{f:scaling_mf}. It turns out that
all these quantities follow power laws: $t_{max} \sim N^{\alpha}$,
$t_{conv} \sim N^{\beta}$, $t_{diff}=(t_{conv}-t_{max}) \sim N^{\delta}$, and
$N_w^{max} \sim N^{\gamma}$ with exponents $\alpha \approx \beta \approx \gamma
\approx \delta \approx 1.5$. More precisely, each agent accumulates
at the peak a number of order $N^{0.5}$ different words, which means
that the necessary memory per agent grows notably when the system size  
is increased.

Let us mention that the values for $\alpha$ and $\gamma$ can in fact 
be understood through simple analytical arguments. Assume indeed that, 
at the maximum,  the average number of words per
agent scales as $c N^a$. The probability for an agent chosen as speaker 
to utter a specific word is $1/(cN^a)$, and the probability that
the hearer already possesses this word is $c N^a/(N/2)$. The 
balance of unsuccessful interactions (which increase $N_w$ by one unit) and
successful ones (which decrease $N_w$ by $2cN^a$) can then be written as:
\begin{equation} 
\frac{dN_w(t)}{dt} \propto \frac{1}{c N^{a}} \left(1- \frac{2 c
N^{a}}{N}\right) - \frac{1}{c N^{a}} \frac{2 c N^{a}}{N} 2 c
N^{a} \ .
\end{equation}
At the maximum, $\frac{dN_w(t)}{dt}=0$, so that the only possibility
is $a=1/2$. Similar arguments can be applied to the derivation of the 
exponent for the time of the peak. It is important to stress that
these analytical results can be obtained thanks to the simplicity of the
microscopic interaction rules.

In summary, the time to convergence grows quite fast as a function of the
system size, and the necessary memory used by each agent also diverges when
$N$ grows. A natural and important question is therefore whether it is
possible to improve the performance of the system. More precisely, a major
challenge would be to improve the population-scale performances of the process
without loosing the simplicity of the microscopic rules, which is the precious
ingredient that allows for in-depth investigations of global-scale dynamics.
We will address this problem in the next section.

\section{Smart Strategy}

In the model described in the previous section, agents, when playing as
speakers, extract randomly a word in their inventories. This feature, along
with the drastic deletion rule that follows a successful game, is the
distinctive trait of the model. Indeed, most of the previously proposed models
of semiotic dynamics prescribe that a weight is associated to each word in
each inventory: this weight determines its probability of being chosen
(see, for instance~\cite{orthogonal}, and references therein). As a natural
consequence the effect of a successful game consists in updating the weights,
rewarding the weight associated with the winning word and possibly reducing the
others. Such sophisticated structures can in principle lead to faster
convergence, but make the models more complicated, compromising the
possibility of a clear global scale picture of the convergence process.

In order to maintain the simplicity of the dynamical rules, it seems natural
to alter the purely stochastic selection rule of the word chosen by the
speaker. In the model previously described, all the words of a given agent's
inventory share a priori the same status. However, a simple parameter to
distinguish between them is their "arrival time", i.e.  the time at which they
enter in his inventory. In particular two words are easily distinguished from
the others: the last recorded one and the last one that gave rise to a
successful game, i.e.  the first that was recorded in the new inventory
generated after the successful interaction. Natural
strategies to investigate consist therefore in choosing systematically
one of these particular words.
We shall refer to these strategies
as "play-last" and "play-first" respectively. Other selection rules are
of course possible but would be either more complicated or more artificial.

The scaling behavior of the model when the "play-last" strategy is adopted is
very interesting (data not shown). The peak time and height scale respectively
as $t_{max} \sim N^{\alpha}$ with $\alpha \approx 1.3$ and $N_w^{max} \sim
N^{\gamma}$ with $\gamma \approx 1.3$, i.e. the used memory is reduced, 
while the convergence time scales as
$t_{conv} \sim N^{\beta}$ with $\beta \approx 2.0$. At the beginning of the
process, playing the last registered word creates a positive feedback that
enhances the probability of a success. In particular a circulating word has
more probabilities of being played than with the usual stochastic rule, thus
creating a scenario in which less circulating words are known by more agents.
On the other hand the "last in first out" approach is highly ineffective when
agents start to win, i.e. after the peak. 
In fact, the scaling $t_{conv} \sim N^{\beta}$ can be explained 
through simple analytical arguments. Let us denote by $N_a$ the number
of agents having the word "a" as last recorded one. This number can
increase by one unit if one of these agents is chosen as speaker,
and one of the other agents is chosen as hearer, i.e. with probability
$N_a/N \times (1-N_a/N)$; the probability to decrease $N_a$ of one
unit is equal to the probability that one of these agents is a hearer
and one of the others is a speaker, i.e. $(1-N_a/N)N_a/N$. These
two probabilities are perfectly balanced so that the resulting process
for the density $\rho_a=N_a/N$ can be written as an unbiased random walk
(with actually a diffusion coefficient $\rho_a (1-\rho_a)/N^2$); it is then
possible to show that the
time necessary for one of the $\rho_a$ to reach $1$ is of order $N^2$.
In summary, in this framework it is
much more difficult to bring to convergence all the agents, since each
residual competing word has a good probability of propagating to other
individuals.

\begin{figure}[t]
\begin{center}
\includegraphics[width=7.5cm]{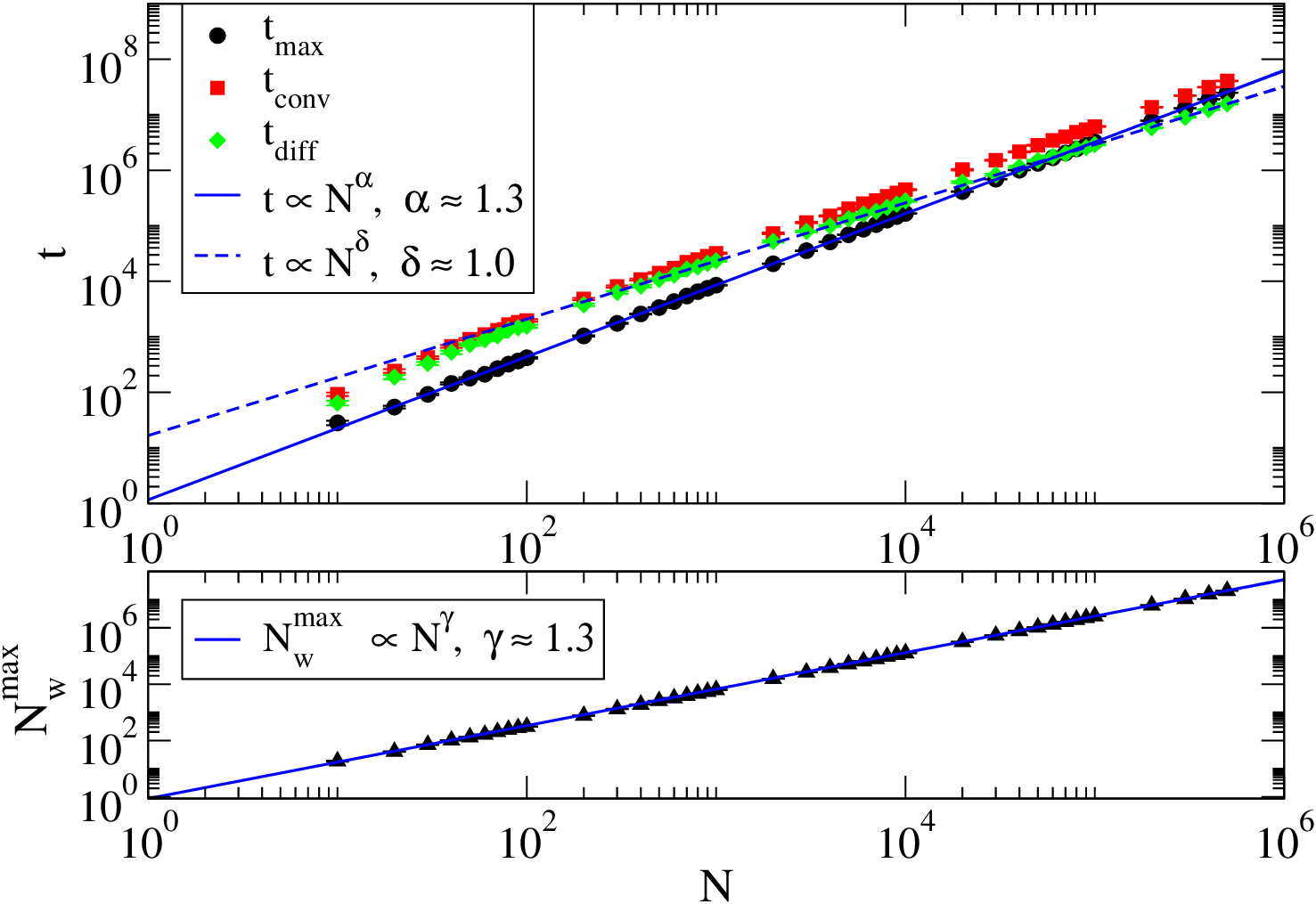}
\caption{Play-smart strategy - scaling with the population size $N$. 
  Top - For the time of the peak $t_{max} \sim N^{\alpha}, \;\; \alpha
  \approx 1.3$, while for the convergence time we have $t_{conv} \sim a
  N^{\alpha} + b N^{\delta}$ with $\delta \approx 1.3, \;\; \delta \approx
  1.0$. Bottom - the maximum number of words scales as $N_w^{max} \sim
  N^{\gamma}$ with $\gamma \approx 1.3$. The "play-smart" rule gives rise to a
  more performing process, from the point of view of both convergence time and
  memory needed. }
\label{f:scaling_ps}
\end{center}
\end{figure}

\begin{figure}[t]
\begin{center}
\includegraphics[width=6.5cm]{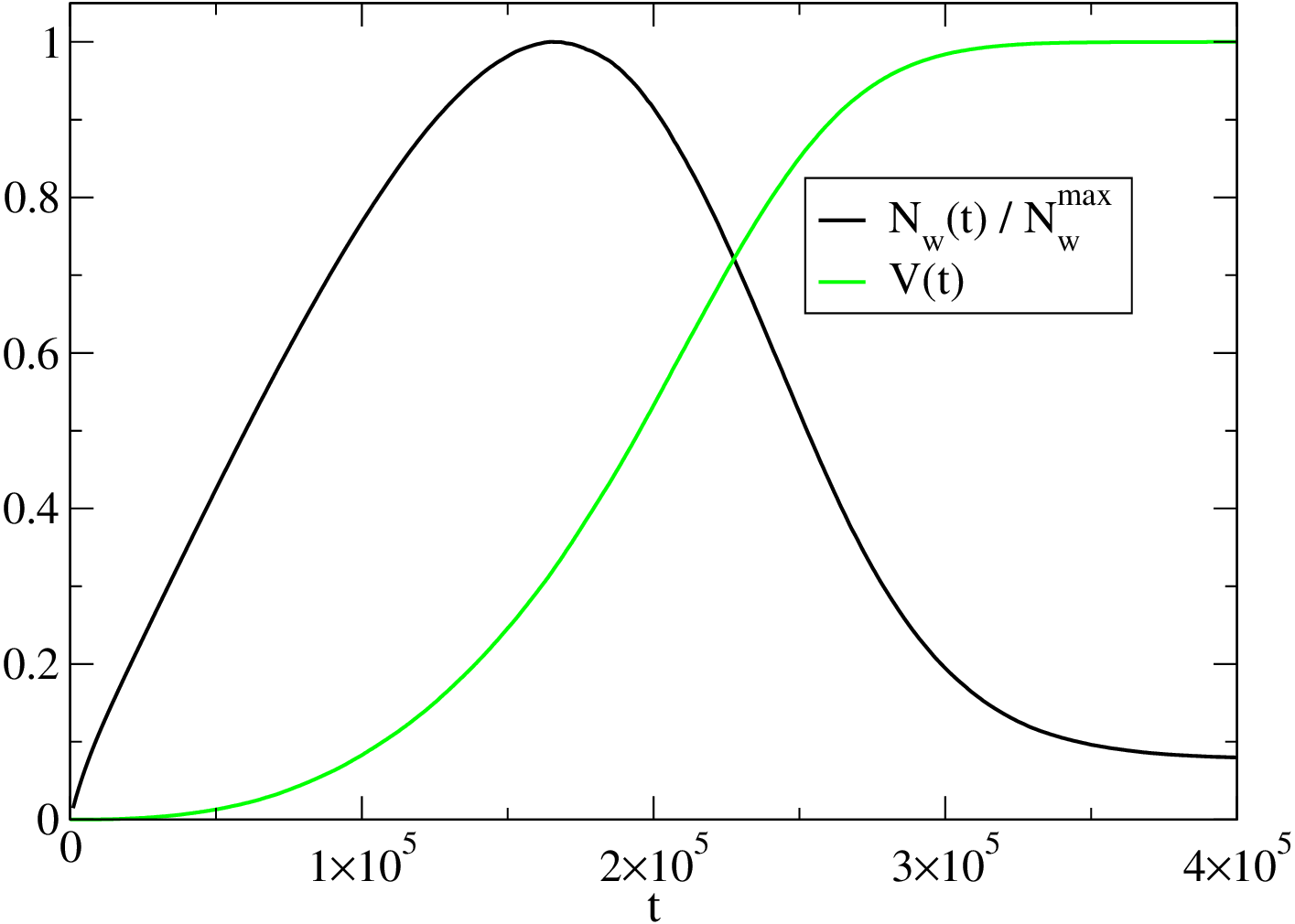}
\caption{"Play-smart" strategy - Fraction $V(t)$ of agents who have 
  played at least one successful game at time t. The transition between the
  initial condition, in which all agents play the last heard word, to the
  final one, in which agents play the word which took part in their last
  successful interaction, is continuous. The growth gets faster after
  the peak of $N_w$.}
\label{f:wintime}
\end{center}
\end{figure}

\begin{figure}[t]
\begin{center}
\includegraphics[width=7.5cm]{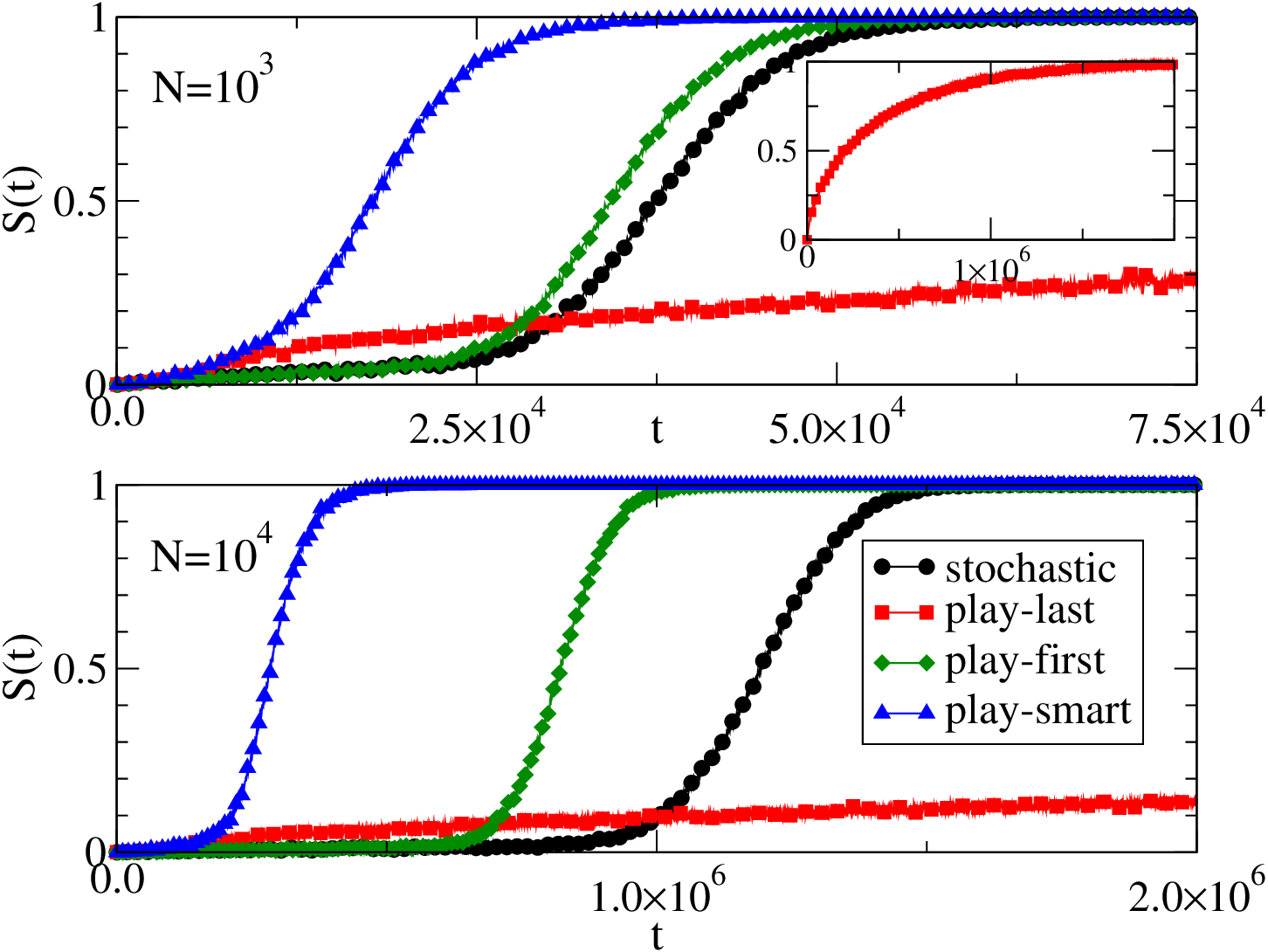}
\caption{Success rate curves $S(t)$ for the various strategies:
  stochastic, "play-last", "play-first" and "play-smart". At 
the beginning of the process the stochastic and "play-first" strategies yield similar success
rates, 
but then the deterministic rule speeds up convergence. On the other hand 
also the "play-smart" and the "play last" evolve similarly at the 
beginning, but the latter reaches the final state much earlier through a 
steep jump. It is worth noting that for three strategies the $S(t)$ 
curves present a characteristic $S-shaped$ behavior, while in 
the "play-last" one the disorder-order transition is more continuous 
(see inset in the top figure). All curves, both for $N=10^3$ 
and $N=10^4$, have been generated averaging over $3\times10^3$ simulation runs.}
\label{f:all_strat_pwin}
\end{center}
\end{figure}

\begin{figure}[t]
\begin{center}
\includegraphics[width=7.5cm]{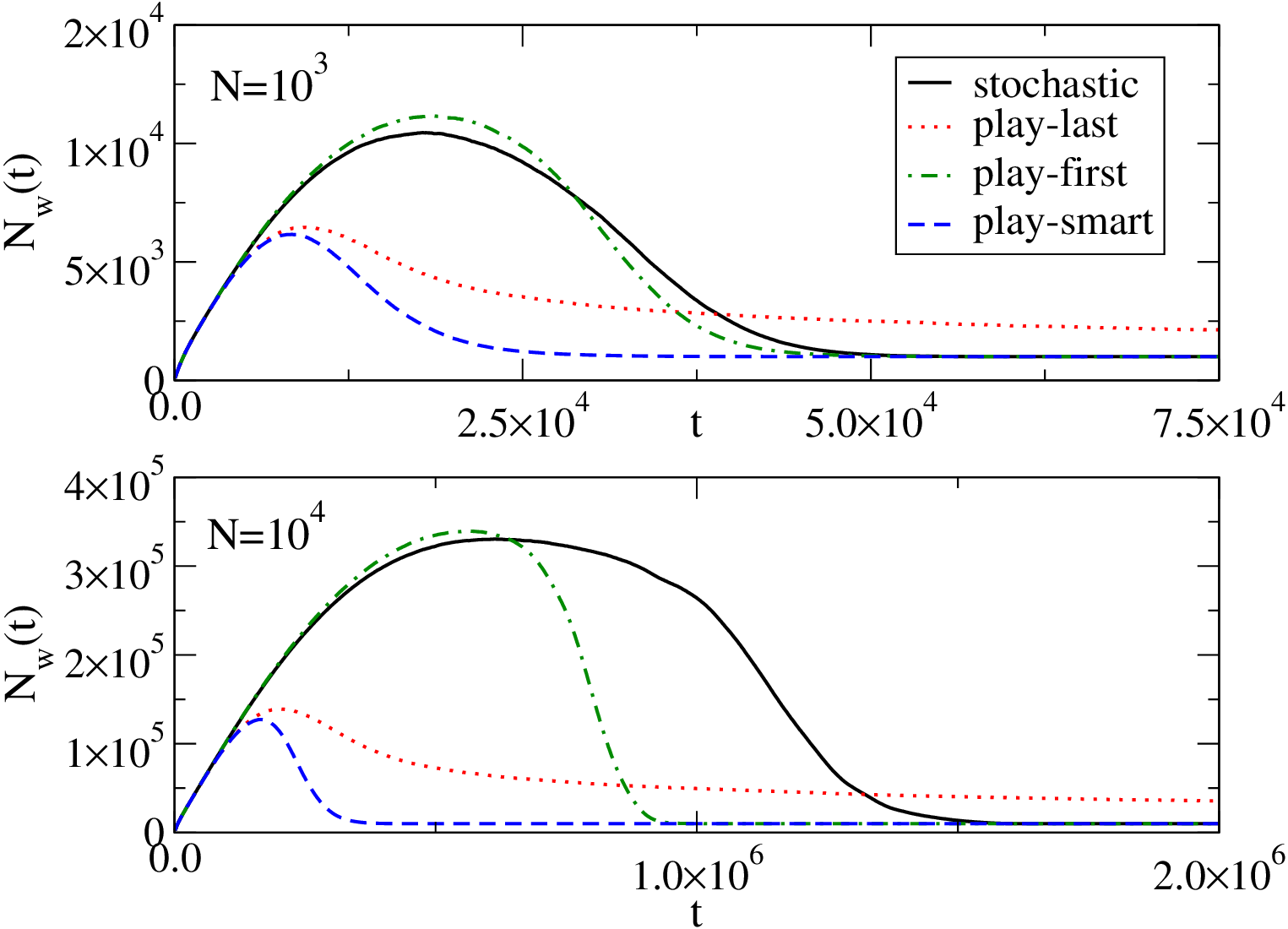}
\caption{Total number of words $N_w(t)$ for the various strategies:
  stochastic, "play-last", "play-first" and "play-smart".  Due to different
  scaling behaviors of the process, differences become more and more relevant
  for larger $N$ (top figure: $N=10^3$; bottom figure: $N=10^4$). The
  "play-smart" approach combines the advantages of "play-last" and
  "play-first" strategies.}
\label{f:all_strat}
\end{center}
\end{figure}

The "play-first" strategy, on the other hand, leads to a faster convergence.
Due to a sort of arbitrariness in the strategy before the first success of the
speaker, the peak related quantities keep scaling as in the usual model, so
that $t_{max} \sim N^{\alpha}$ and $ N_w^{max} \sim N^{\gamma}$ with $\alpha
\approx \gamma \approx 1.5$ (data not shown). This seems natural, since
playing the first recorded word is essentially the same as extracting it
randomly when most agents have only few words. In fact, in both cases no
virtuous correlations or feedbacks are introduced between circulating and
played words. However, playing the last word which gave rise to a successful
interaction strongly improves the system-scale performances once the agents
start to win. In particular it turns out that for the difference between the
peak and convergence time we obtain $(t_{conv}-t_{max}) \sim N^{\delta}$ with
$\delta \approx 1.15$ (data not shown), so that the behavior of the
convergence time is the result of the combination of two different power law
regimes, i.e.  $t_{conv} \sim a N^{\alpha} + b N^{\delta}$.  On the other
hand, the stochastic rule leads to $(t_{conv}-t_{max}) \sim N^{1.5}$ as shown
in Figure~\ref{f:scaling_mf}.  This means that the "play-first" strategy is
able to reduce the time that the system has to wait before reaching the
convergence, after the peak region. This seems the natural consequence of the
fact that successful words increase their chances to be played while
suppressing the spreading of other competitors.

In summary, we have seen that, compared to the usual random extraction of the
played word, the "play-last" strategy is more performing at the beginning of
the process, while the "play-first" one allows to fasten the convergence of
the process, even if it is effective only after the peak of the total number
of words. It seems profitable, then, to define a third alternative strategy
which results from the combination of the two we have just described. The new
prescription, which we shall call "play-smart", is the following:
 
\begin{itemize}
\item[$\rightarrow$] If the speaker has never took part in a successful game,
  he plays the last word recorded;
\item[$\rightarrow$] Else, if the speaker has won at least once, he plays the
  last word he had a communicative success with.
\end{itemize}

The first rule will thus be applied mostly at the beginning, and as the system
evolves, the second rule will be progressively adopted by more and more
agents. Since the change in strategy is not imposed at a given time, but
takes place gradually, in a way depending of the evolution of the system,
such a strategy has also the interest of being in some sense 
self-adapting to the system's actual state.
In Figure~\ref{f:scaling_ps}, the scaling behaviors relative to the
"play-smart" strategy are reported. Both the height and time of the maximum
follow the scaling of the "play-last" strategy: $t_{max} \sim N^{\alpha}$
and $N_w^{max} \sim N^{\gamma}$ with $\alpha \approx \gamma \approx 1.3$. The
convergence time, on the other hand, scales as a superposition of two power
laws: $t_{conv} \sim a N^{\alpha} + b N^{\delta}$ with $\alpha \approx 1.3,
\delta \approx 1.0$. Thus, the global behavior determined by the
"play-smart" modification is indeed less demanding in terms of both memory and
time. In particular, while the lowering of the peak height yields
in fact a slower convergence for the "play-last" strategy, the progressive
self-driven change in strategy allows to fasten the convergence
further than for the "play-first" strategy.

It is also worth studying how the transition between the initial situation in
which most agents play the last recorded word to that in which they play the
last successful word takes place in the "play-smart" strategy. In other words
we want to study the probability $V(t)$ of finding an agent that has already
been successful in at least one interaction at a given time. Results relative
to a population of $10^4$ agents are shown in Figure~\ref{f:wintime}.
Interestingly, the transition from the initial situation to the final one is
continuous, and there is a sudden speeding up after the peak. 

Finally, in order to have an immediate feeling of what different playing word
selection strategies imply, we report in Figures~\ref{f:all_strat_pwin}
and~\ref{f:all_strat} the success rate $S(t)$ and the total
number of words, $N_w(t)$ relative to the four strategies described
previously, for two different sizes. 
The "play-first" and "play-smart" curves exhibit the same 
"S-shaped" behavior for $S(t)$ as in the case of the stochastic model, while
the "play-last" rule affects qualitatively 
the way in which the final state is reached. Indeed, in this case the transition 
between the initial disordered state and the final ordered one
is more continuous 
(see the inset in the top figure).
Moreover, Figure~\ref{f:all_strat} illustrates 
that the choice of the strategy has substantial quantitative
consequences for both necessary memory and
time needed to reach convergence, even if
the changes in scaling behavior could at first appear rather
limited (from $N^{1.5}$ to $N^{1.3}$).
In particular, the "play-smart" strategy, which adapts
itself to the state of the system, leads to a drastic reduction
of the memory and time costs and thus to a dramatic increase in efficiency.

\section{Conclusion}

In conclusion, we have discussed a multi-agent model of Semiotic Dynamics
which is able to describe the convergence of a population of agents on the use
of a particular semiotic convention (a name to assign to an object, in our
case). The model relies on very simple microscopic interaction rules, thus
being appropriate for accurate global scale investigations. We have then
shown that the modification of the rule followed by agents to select the word
to be transmitted gives rise to a process which is less demanding in terms of
agents memory usage and leads to a faster convergence, too. Due to the
possible utility of Semiotic Dynamics models for the design of technological
systems, we believe that the findings presented here are not only
intrinsically interesting from a theoretical point of view, but can also be
relevant for applications.

\section{Acknowledgements}
The authors thank L. Steels and C. Cattuto for many stimulating discussions.
A. Baronchelli and V. L. have been partly supported by the ECAgents project
funded by the Future and Emerging Technologies program (IST-FET) of the
European Commission under the EU RD contract IST-1940. The information
provided is the sole responsibility of the authors and does not reflect the
Commission's opinion. The Commission is not responsible for any use that may
be made of data appearing in this publication.  A. Barrat and L.D. are
partially supported by the EU under contract 001907 (DELIS).

\bibliography{alife_ng_bib}

\begin{thebibliography}{}

\bibitem[Axelrod, 1997]{Axelrod:1997}
Axelrod, R. (1997).
\newblock The dissemination of culture: A model with local convergence and
  global polarization.
\newblock {\em Journal of Conflict Resolution}, 41:203--226.

\bibitem[Baronchelli et~al., 2005]{ng_first}
Baronchelli, A., Felici, M., Caglioti, E., Loreto, V., and Steels, L. (2005).
\newblock Sharp transition towards shared vocabularies in multi-agent systems.
\newblock {\em arxiv:physics/0509075}, submitted for publication.

\bibitem[Deffuant et~al., 2000]{Deffuant:2000}
Deffuant, G., Neau, D., Amblard, F., and Weisbuch, G. (2000).
\newblock Mixing beliefs among interacting agents.
\newblock {\em Adv. Compl. Syst.}, 3:87--98.

\bibitem[Hurford, 1989]{hurford89}
Hurford, J. (1989).
\newblock Biological evolution of the saussurean sign as a component of the
  language acquisition device.
\newblock {\em Lingua}, 77(2):187--222.

\bibitem[Krapivsky, 1992]{Krapivsky:1992}
Krapivsky, P. (1992).
\newblock Kinetics of monomer-monomer surface catalytic reactions.
\newblock {\em Phys. Rev. A}, 45:1067--1072.

\bibitem[Lenaerts et~al., 2005]{orthogonal}
Lenaerts, T., Jansen, B., Tuyls, K., and Vylder, B.~D. (2005).
\newblock The evolutionary language game: An orthogonal approach.
\newblock {\em Journal of Theoretical Biology}, 235:566--582.

\bibitem[Nowak et~al., 1999]{nowak99}
Nowak, M.~A., Plotkin, J.~B., and Krakauer, D. (1999).
\newblock The evolutionary language game.
\newblock {\em Journal of Theoretical Biology}, 200(2):147--162.

\bibitem[Steels, 1995]{ng1995}
Steels, L. (1995).
\newblock A self-organizing spatial vocabulary.
\newblock {\em Artificial Life}, 2(3):319--332.

\bibitem[Steels, 1999]{talking_heads}
Steels, L. (1999).
\newblock {\em The Talking Heads Experiment. Volume 1. Words and Meanings}.
\newblock Laboratorium, Antwerpen.

\bibitem[Steels, 2000]{steels_dynsys}
Steels, L. (2000).
\newblock Language as a complex adaptive system.
\newblock In Schoenauer, M., editor, {\em Proceedings of PPSN VI}, Lecture
  Notes in Computer Science, Berlin, Germany. Springer-Verlag.

\bibitem[Sznajd-Weron and Sznajd, 2000]{Sznajd:2000}
Sznajd-Weron, K. and Sznajd, J. (2000).
\newblock Opinion evolution in closed community.
\newblock {\em Int. J. Mod. Phys. C}, 11:1157--1165.

\end{thebibliography}
\bibliographystyle{alife10}

\end{document}